\pdfoutput=1
\documentclass[10pt,preprint]{aastex}
\usepackage{a4wide,multicol,graphicx}
\newif\ifAMStwofonts
\AMStwofontstrue

\def\ts   {\thinspace}

\newcommand\HII{\mbox{H\thinspace{\sc ii}}}
\def\hi  {\ifmmode{{\rm H}{\rm \small I}}\else{H\ts {\scriptsize I}}\fi}
\def\Ha  {\ifmmode{{\rm H}{\alpha}}\else{H\ts {$\alpha$}}\fi}
\newcommand{\D}{$^\circ$}
\def\arcmin{\hbox{$^{\prime}$}}
\def\arcsec{\hbox{$^{\prime\prime}$}}
\def\aco {\ifmmode{^{12}{\rm CO}(J=1\to0)}\else{$^{12}{\rm
CO}(J=1\to0)$}\fi}
\def\msol   {\ifmmode{{\rm M}_{\odot}}\else{M$_{\odot}$}\fi}
\begin{document}

\title{An ATCA 20cm Radio Continuum Study of the Large Magellanic
  Cloud}\author{A. Hughes$^{1,2}$\thanks{Email:
    ahughes@astro.swin.edu.au}, L. Staveley-Smith$^{3}$, S. Kim$^{4}$,
  M. Wolleben$^{5,6}$, M. Filipovi\'c$^{7}$ \\ $^1$ Centre for
  Supercomputing and Astrophysics, Swinburne University of Technology,
  Hawthorn VIC 3122, Australia \\ $^2$ CSIRO Australia Telescope
  National Facility, PO Box 76, Epping NSW 1710, Australia \\ $^3$
  School of Physics M013, University of Western Australia, Crawley WA
  6009, Australia \\ $^4$ Department of Astronomy and Space Science,
  Sejong University, KwangJin-gu, KunJa-dong 98, Seoul 143-747, Korea
  \\ $^{5}$ Department of Electrical and Computer Engineering,
  University of Alberta, Edmonton, AB T6G 2V4, Canada \\ $^{6}$
  National Research Council Canada, Herzberg Insitute of Astrophysics,
  Dominion Radio Astrophysical Observatory,\\ Penticton, BV V2A 6J9,
  Canada \\ $^7$ University of Western Sydney, Locked Bag 1797, Penrith
  South, DC, NSW 1797, Australia \\ }

\date{Typeset \today; Received / Accepted}

\maketitle

\label{firstpage}

\begin{abstract}
\noindent We present a mosaic image of the 1.4~GHz radio continuum
emission from the Large Magellanic Cloud (LMC) observed with the
Australia Telescope Compact Array (ATCA) and the Parkes Telescope. The
mosaic covers 10.8\D $\times$ 12.3\D\ with an angular resolution of
40\arcsec, corresponding to a spatial scale of $\sim10$~pc in the
LMC. The final image is suitable for studying emission on all scales
between 40\arcsec\ and the surveyed area. In this paper, we discuss i)
the characteristics of the LMC's diffuse and compact radio continuum
emission, ii) the fraction of the emission produced by thermal
processes and the implied star formation rate in the LMC, and iii)
variations in the radio spectral index across the LMC. Two
non-standard reduction techniques that we used to process the ATCA
visibility data may be of interest for future wide-field radio
continuum surveys. The data are open to the astronomical community and
should be a rich resource for studies of individual objects such as
supernova remnants (SNRs), \HII\ regions and planetary nebulae (PNe),
as well as extended features such as the diffuse emission from
synchrotron radiation.
\end{abstract}


\section{INTRODUCTION}
\label{sect:introduction}
\noindent Radio continuum emission is a useful tool for studying star
formation processes in galaxies. The two main components of the
emission are thermal free-free radiation from ionized gas in \HII\
regions and synchrotron radiation emitted by relativistic electrons
accelerated in magnetic fields \citep{condon92}. Both processes are
thought to be related to the evolution of massive stars, but they
provide information about different periods in the galaxy's star
formation history. Thermal radio emission arises directly from the
ionized gas surrounding young massive stars. The intensity of the
emission is proportional to the total number of Lyman continuum
photons, and in the optically thin regime the spectrum is nearly flat
($\alpha=-0.1$, where we adopt the convention $S_{\nu} \propto
\nu^{\alpha}$). For an isolated star-forming region, the thermal radio
emission should persist over time-scales similar to the average
lifetime of an \HII\ region ($\sim$10~Myr), suggesting that the thermal
component of a galaxy's radio continuum emission should be a good
tracer of the current star formation rate
\citep[e.g.][]{kennicutt98}. The nonthermal emission, on the other
hand, originates in the supernova explosions and supernova remnants
(SNRs) that occur at the end of a massive star's life. If discrete
SNRs were solely responsible for accelerating the relativistic
electrons that emit synchrotron radiation, then the time-scale of the
nonthermal radio emission would also be relatively short and the
emission would provide another tracer of the galaxy's current star
formation activity. However, observations of normal spiral galaxies
suggest that only $\sim$10\% of a galaxy's synchrotron emission is due
to electrons accelerated in the magnetic field of discrete SNRs; the
remainder is from electrons accelerated in the widespread galactic
field over time-scales of 10 to 100~Myr \citep{heloubicay93}. For a
galaxy with a single isolated star forming region, the emission from
discrete SNRs might still be expected to dominate the nonthermal
component at early times (before the relativistic electrons have had
time to diffuse away from their production sites), while very young
starbursts ($\leq$3~Myr) should show almost no synchrotron radiation
since the massive stars have not yet evolved into supernovae. These
general considerations about the time-scales of the thermal and
nonthermal radio continuum emission have prompted several groups to
explore whether the radio spectral index might prove to be a useful
method to chronicle the star formation activity of starburst activity
in simple systems
\citep[e.g][]{bressanetal02,cannonskillman04,hirashitahunt06}.\\

\noindent In this paper, we present a new, high-resolution survey of
the 1.4~GHz radio continuum emission from the Large Magellanic Cloud
(LMC). The LMC is a gas-rich, irregular dwarf galaxy that exhibits
clear signs of active star formation. Reddening and extinction due to
dust in the LMC are low \citep[$E_{B-V}\sim0.13$,][]{masseyetal95},
and the LMC's declination is such that our view of the galaxy is
mostly uncontaminated by foreground emission from the Milky Way. The
inclination of the LMC is also reasonably slight
\citep[$i\sim35$\D,][]{vandermarelcioni01}, minimizing line-of-sight
confusion. The LMC thus presents a unique opportunity to study an
entire galaxy at high angular resolution. At an assumed distance of
50.1~kpc \citep{alves04}, 1 \arcmin\ corresponds to 15~pc, making the
LMC an excellent laboratory to investigate the relationship between
different phases of the interstellar medium (ISM), the interaction of
the ISM with individual objects, and the influence of galactic-scale
processes on the properties of interstellar material. Recently
completed surveys such as the Magellanic Cloud Emission Line Survey
\citep[MCELS,][]{smithetal98}, the ATCA+Parkes HI 21cm survey
\citep{kimetal98,kimetal03}, the Spitzer SAGE project in the
far-infrared \citep{meixneretal06} and the \aco\ survey by NANTEN
\citep{fukuietal01} have the potential to provide a comprehensive view
of the dust and gas phases in the LMC's interstellar medium, along
with a complete inventory of stars and proto-stars. A high angular
resolution survey of the radio continuum emission in the LMC is a
timely and important complement to these datasets. In addition to the
total intensity images that we present here, new studies of the
polarisation and Faraday rotation of background sources in the
ATCA+Parkes radio continuum data have begun to reveal the strength and
detailed structure of the LMC's magnetic field
\citep{gaensleretal05b,gaensleretal05}.\\

\noindent This paper is organized as follows. In Section
\ref{sect:observations} we present the observing strategy of our
survey. The methods that we used to reduce and combine the
interferometer and single-dish data are described in Section
\ref{sect:data}. In Section \ref{sect:results}, we present the final
1.4~GHz image of the LMC and examine spatial variations in the LMC's
radio spectral index. We estimate the thermal fraction of the radio
continuum emission, and compare the radio-derived star formation rate
for the LMC to estimates determined via other star formation rate
calibrations. Section \ref{sect:summary} contains a summary of our
conclusions and outlines some potential future uses for the radio
continuum data.

\section{OBSERVATIONS}
\label{sect:observations}

\noindent These observations of the LMC were conducted at the
Australia Telescope Compact Array (ATCA) at the time of the HI survey
by \citet{kimetal98,kimetal03} in a second IF band centred on
1384~MHz. The ATCA is an east-west interferometer located at the Paul
Wild Observatory in Narrabri, Australia. The latitude of the ATCA is
-30\D18\arcmin. The interferometer consists of five 22~m antennae
positioned along a 3~km track, with a sixth antenna located 3~km from
the western end of the track. Observations of the LMC were made with
four 750~m arrays between October 1994 and February 1996. The
observing log is presented in Table~\ref{tbl:obslog}. Across the four array
configurations, there are a total of 40 independent baselines ranging
from 30 to 750~m. The ATCA antenna stations are regularly spaced, with
the consequence that all possible baselines are incremented by
multiples of 15.3~m.\\

\begin{table}
\caption{Summary of observing dates and array configurations.}
\label{tbl:obslog}
\begin{center}
\begin{tabular}{l|r|}
\hline
Date & Array Configuration \\
\hline
1994 Oct 26 - Nov 9 & 750D \\
1995 Feb 23 - Mar 11 & 750A \\
1995 Jun 02 - Jun 07 & 750B \\
1995 Oct 15 - Oct 31 & 750B \\
1996 Jan 27 - Feb 8 & 750C \\
\hline
\end{tabular}
\end{center}
\end{table}

\begin{figure}
\begin{center}
\includegraphics[width=12cm]{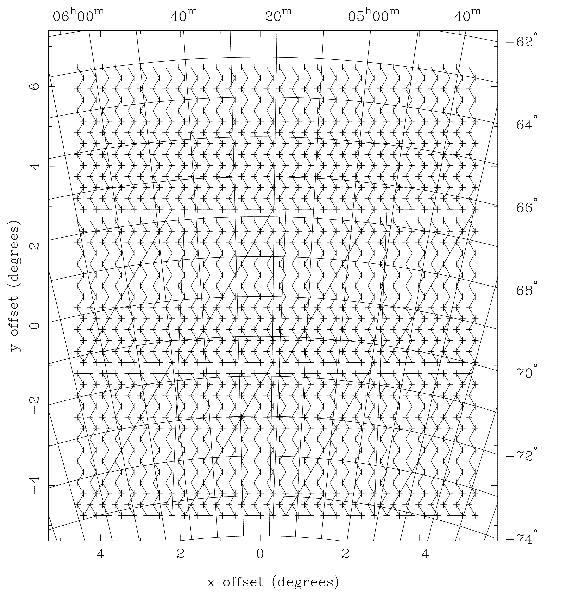}
\caption{The scanning strategy and individual pointing centres of the
  ATCA LMC mosaic. }
\label{fig:scan}
\end{center}
\end{figure}

\begin{figure}
\begin{center}
\includegraphics[width=10cm,angle=270,clip]{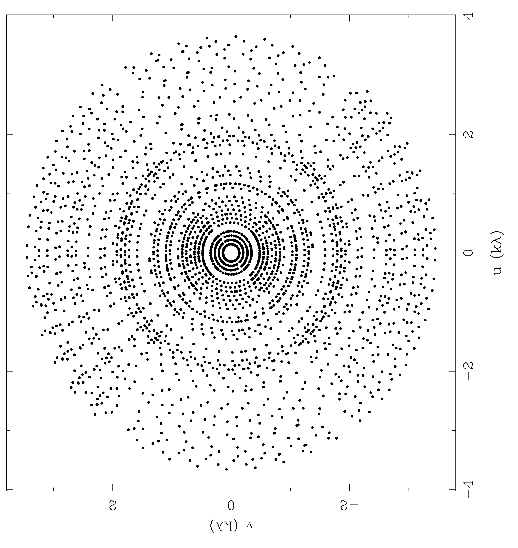}
\caption{$u-v$ coverage of a single pointing within the ATCA LMC
  mosaic.}
\label{fig:pctrs}
\end{center}
\end{figure}

\noindent For our survey, we mapped a 10.8\D $\times$ 12.3\D\ field
covering the LMC, centred on (05h20m,-68d44m)$_{J2000}$. We
divided the total survey area into 12 regions, each containing 112
pointing centres. The array was cycled around the 112 pointing centres
within each region according to a hexagonal grid pattern determined by
Nyquist's theorem. In this case, the angular separation of the
pointing centres is given by
\begin{equation}
\theta=\frac{2}{\sqrt{3}}\frac{\lambda}{2D}
\end{equation}
where $\lambda$ is observing wavelength and $D$ is the diameter of the
antenna. For our observations, $\lambda = 21$~cm and $D = 22$~m,
giving a pointing centre separation of $\theta=19$ arcmin. Each
pointing was observed between 95 and 140 times during the entire
survey, which corresponds to between 18 and 26 minutes of total
integration time per pointing. A map of the pointing centres and
scanning direction for the ATCA mosaic is shown in
Fig.~\ref{fig:scan}. The $u-v$ coverage for a single pointing centre
within the mosaic is shown in Fig.~\ref{fig:pctrs}.\\

\noindent All our observations were recorded in wideband continuum
mode with 32 4~MHz channels across a total bandwidth of 128~MHz. The
centre frequency was 1.384~GHz. The ATCA feeds receive two orthogonal
linear polarisations, X and Y. For the continuum observations, we
measured the four polarisation products XX, YY, XY, and YX. Here we
only discuss the total intensity data; a preliminary analysis of the
polarised emission has already been presented
\citep{gaensleretal05b}. The \hi\ emission from neutral hydrogen in
the LMC was recorded simultaneously in the second frequency chain. The
processing and analysis of the \hi\ data have been described elsewhere
\citep{kimetal98,kimetal03}. \\

\noindent Previous studies that have made use of the radio continuum
data that we present here include \citet{cohenetal03} and
\citet{hughesetal06}. The data used in these publications were
processed according to the procedure outlined in
Section~\ref{sect:data}, except that the peeling technique (described
in Section~\ref{sect:peeling}) was not applied.\\

\section{DATA REDUCTION}
\label{sect:data}
\noindent The ATCA data were flagged, calibrated and imaged using the
\textsc{MIRIAD} software package \citep{saultetal85}. We used the
source PKS B1934-638 for bandpass and absolute flux density
calibration (the flux density of PKS B1934-638 at 1.377~GHz is
14.95~Jy). One of either PKS B0407-658 or PKS B0454-810 was observed
every 30 minutes in order to calibrate the time variation in the
complex antenna gains. PKS B1934-638 has no detectable linear
polarization and can thus be used to solve for polarization
leakages. We had sufficient parallactic angle coverage of the two
secondary calibrators to disentangle their intrinsic and instrumental
polarization, allowing us to calibrate Stokes U and Q as well as total
intensity.\\

\noindent The individual pointings were linearly combined and imaged
using a standard grid-and-FFT scheme with superuniform weighting. Like
uniform weighting, superuniform weighting minimizes sidelobe levels to
improve the dynamic range and sensitivity to extended structure of the
final mosaicked image. Uniform weighting reduces to natural weighting,
however, if the total field-of-view of the mosaic is much larger than
the primary beam. Superuniform weighting overcomes this limitation by
decoupling the weighting from the size of the field. It attempts to
minimize sidelobe contributions from strong sources over a region
smaller than the total image and is typically more successful than
uniform weighting for large mosaics \citep{saultetal96}.

\subsection{Image Deconvolution}
\label{sect:deconv}
\noindent We developed a two-step Fourier deconvolution strategy for
the our LMC data. After inverting the mosaic visibilities, we
constructed a preliminary CLEAN model of our data by using 1.2 million
iterations of the Steer-Dewdney-Ito (SDI) CLEAN algorithm on our dirty
map \citep{steeretal84}. The residuals of the CLEAN model, mainly
corresponding to diffuse emission, were deconvolved using maximum
entropy. The CLEAN model and the maximum entropy model were linearly
combined and restored with a 40\arcsec\ Gaussian beam in order to form
the final image.

\subsection{Peeling}
\label{sect:peeling}
\noindent The deconvolved ATCA image exhibits ring-like artefacts at
the $\sim$0.5\% level. These become significant close to bright
compact sources such as 30 Doradus, limiting the sensitivity that can
be achieved in these regions. These artefacts are mainly due to errors
in the calibration of off-axis sources. There are number of possible
causes for off-axis calibration errors, including pointing errors,
small differences between individual antenna dishes, errors in the
primary beam model, and the rotation of the primary beam diffraction
lobes through off-axis sources. In order to improve the dynamic range
of the ATCA image, we applied a ``peeling'' technique that has been
described by Tom Oosterloo ({\it priv. comm.}). Contrary to the
usual assumption that one set of antenna gain solutions is adequate
across the field of a single pointing, peeling explicitly solves for
the antenna gains at the position of off-axis sources, so that
different calibration solutions can be applied to different regions
within the field of each pointing. \\

\begin{figure*}
\includegraphics[width=7cm,angle=0]{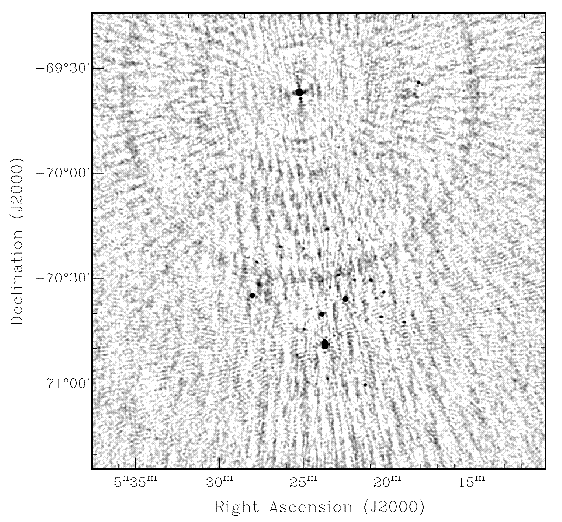}
\includegraphics[width=7cm,angle=0]{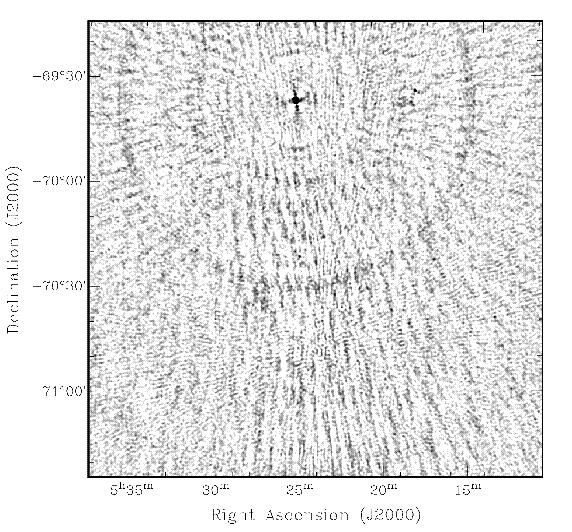}
\includegraphics[width=7cm,angle=0]{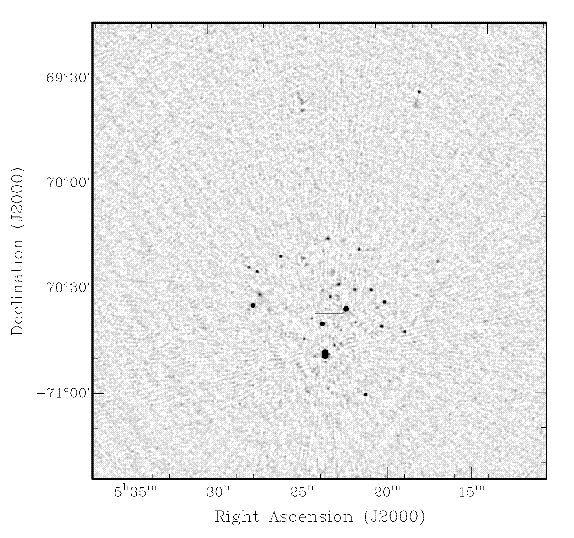}\hfill
\includegraphics[width=7cm,angle=0]{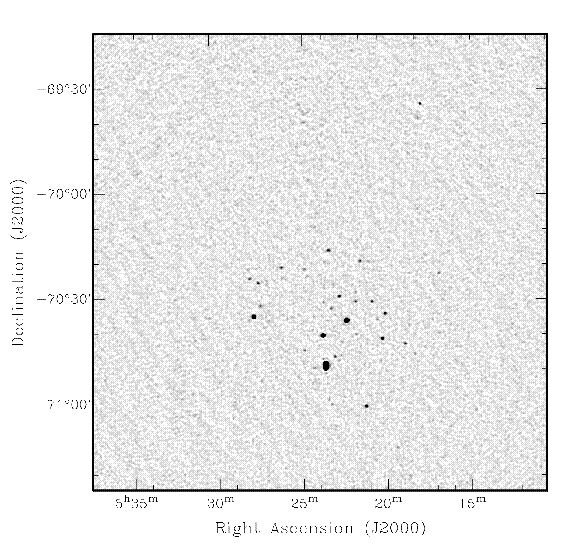}
\caption{Example of peeling procedure for a single pointing of the
  ATCA mosaic. Only the region of sky containing the field centre and
  off-axis source is shown. The field centre is
  indicated with a black cross. {\it Top Left.} Original image of the
  pointing. {\it Top Right.}  Image of the pointing after the on-axis
  sources have been subtracted. The appropriate gain solution for the
  off-axis source is then determined from this data using
  self-calibration. {\it Bottom Left.}  Image of the pointing after
  subtracting the off-axis source and its associated errors.  {\it
    Bottom Right.}  Final image of the pointing after performing
  self-calibration on the on-axis sources.}
\label{fig:lmc0812}
\end{figure*}

\noindent In order to determine which pointings were badly affected by
errors from off-axis sources, we deconvolved and imaged the visibility
data for each of the 1344 pointings. Peeling was attempted if the
following criteria were satisfied: i) the flux density of the off-axis
source was greater than 10~mJy~beam$^{-1}$ ii) the off-axis source was
located more than 1.5 times the primary beam FWHM from the pointing
centre and iii) errors due to the off-axis source were evident within
the primary beam. If these criteria were not satisfied, no corrections
to the visibility data were made. If off-axis errors were deemed
significant, we constructed a simple model of the on-axis sources
using 10000 iterations of the SDI CLEAN algorithm, and subtracted the
model of the on-axis sources from the visibility data. The resulting
visibility data (the ``model data'') represent the off-axis source and
its associated errors. The model data were imaged, and a model of the
off-axis source was constructed using 10000 iterations of the SDI
CLEAN algorithm. We then performed an amplitude and phase
self-calibration on the model data in order to obtain a good set of
antenna gain solutions for the off-axis source (the ``model
gains''). The model gains were applied to the original visibility data
for the pointing, and then the model of the off-axis source was
subtracted. Next, the model gains were ``un-applied'' to the
model-subtracted visibility data, i.e. having multiplied the original
visibility data by the antenna gain solutions for the off-axis source,
we multiplied the model-subtracted visibility data by the inverse of
these model gains. At the end of this process, we are left with
visibility data that are identical to the original visibility data for
the pointing, except that the off-axis source and its errors have been
removed. As an example, Fig.~\ref{fig:lmc0812} illustrates the main
stages of the peeling process for a single pointing in the ATCA
mosaic.\\

\begin{figure*}
\begin{center}
\includegraphics[width=15cm,angle=270]{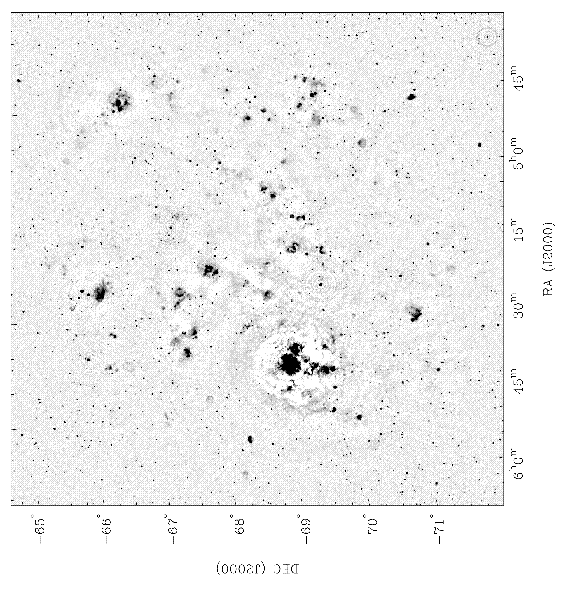}
\caption{ The deconvolved ATCA 1.4~GHz mosaic, after peeling and
  self-calibration of individual pointings but prior to combination
  with the single-dish data. }
\label{fig:ATCAmap}
\end{center}
\end{figure*}

\begin{figure}
\begin{center}
\includegraphics[width=8cm,angle=0]{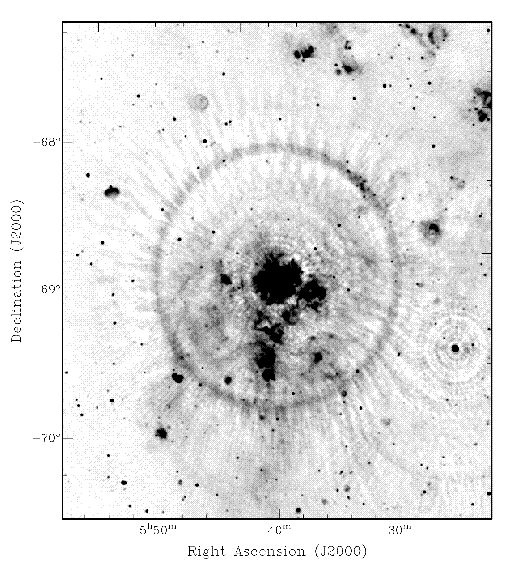}
\includegraphics[width=8cm,angle=0]{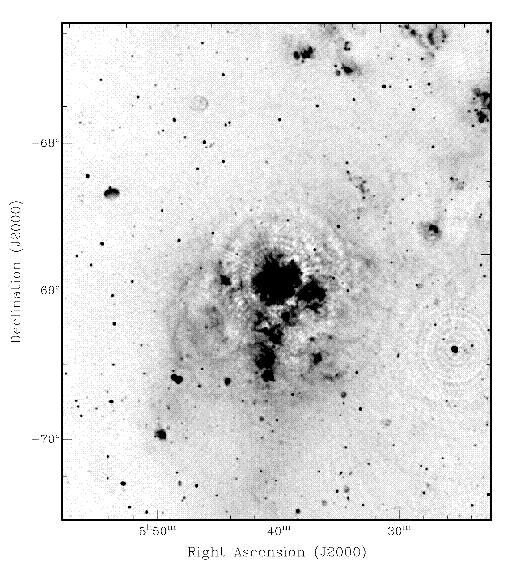}
\caption{1.4~GHz emission in the 30 Doradus region of the LMC. The two
  panels illustrate the final ATCA+Parkes image of this region with
  ({\it top}) and without ({\it bottom}) applying peeling corrections
  to the visibility data.}
\label{fig:ring}
\end{center}
\end{figure}

\noindent For a number of pointings, additional corrections to the
basic antenna gain solutions were required due to calibration errors
for sources located within the primary beam. To improve the antenna
gain solutions for these pointings, amplitude and phase
self-calibration was applied. In total, the peeling technique was
applied to 269 of the 1344 pointings. On-axis self-calibration was
applied to a further 78 pointings. The final set of corrected
visibility data were combined, deconvolved, imaged according to the
strategy described in Section~\ref{sect:deconv} above. The final
ATCA-only mosaic is shown in Fig.~\ref{fig:ATCAmap}. To highlight the
improvement achieved by applying the peeling process, we present a
expanded view of the 30 Doradus region in Fig.~\ref{fig:ring}. 

\subsection{Combination of Inteferometer and Single Dish Data}
\label{sect:combination}
\noindent Although mosaicing recovers angular scales larger than
normal interferometric observations by reducing the effective shortest
projected baseline, our observations are still limited to angular
scales smaller than $\theta = \lambda/(d - D/2)$, where $d=30.6$~m is
the shortest baseline of the ATCA, and $D=22$~m is the diameter of a
single antenna. In our case, the ATCA data is limited to angular
scales smaller than $\sim34$ arcmin. To recover information on larger
scales, the ATCA mosaic data was combined with single-dish data from
the Parkes Telescope. Here we only provide information about the
Parkes data that is relevant for the combination process; a detailed
description of the single-dish observations and data reduction was
presented in \citet{haynesetal86}.\\

\noindent Single-dish and interferometer data may be combined in the
Fourier domain after deconvolution of the individual pointings or in
the $u-v$ plane prior to deconvolution. \citet{stanimirovic02} showed
that comparable results are achieved using either method but combining
the data after deconvolution typically produced results that were more
consistent with other methods. We chose to combine the data in the
Fourier plane after deconvolution. In this method, the ATCA continuum
data are imaged and deconvolved, the single-dish data are imaged, and
the clean interferometric and single-dish images are then Fourier
transformed and combined. The method is implemented in the
\textsc{MIRIAD} task {\tt immerge}. \\

\noindent Slight differences in calibration of the interferometer and
single-dish data can necessitate a relative flux correction
factor. This correction factor is determined by comparing the data
sets in the Fourier plane at every pixel and frequency in the range of
overlapping spatial frequencies. To calculate the calibration factor,
both images must be deconvolved, a step which requires a good
knowledge of the single-dish beam. Using a two-dimensional Gaussian
with FWHM=16.6\arcmin\ for the Parkes beam and by comparing the flux
densities of strong, compact sources in the Parkes and ATCA data, we
calculated a relative calibration factor of 1.07. The final
ATCA+Parkes combined image is shown in Fig.~\ref{fig:map}. 

\subsection{Characteristics of the Final Image}
\label{sect:characteristics}
\begin{figure*}
\begin{center}
\includegraphics[width=15cm,angle=270]{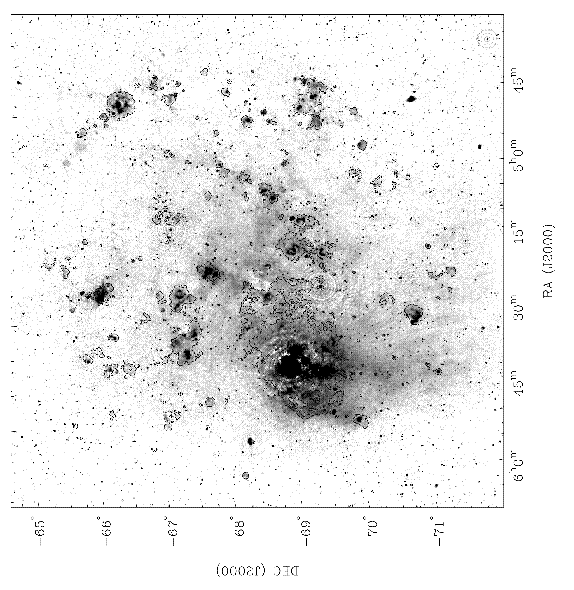}
\caption{The final combined Parkes + ATCA 1.4~GHz radio continuum map
  of the LMC. A square-root intensity scale has been used to emphasise
  the characteristics of the diffuse emission. The black contours
  indicate an \Ha\ brightness of 50~R (compare Fig.~\ref{fig:shassa}).}
\label{fig:map}
\end{center}
\end{figure*}

\noindent The final combined image shown in Fig.~\ref{fig:map} is
sensitive to all angular scales from the synthesized beam size (40\arcsec)
up to the final image size. In order to estimate the map sensitivity,
we measured the average rms of blank regions of sky, finding 0.3 mJy
per 40\arcsec\ beam for the ATCA data, 30 mJy per 16\arcmin.6 beam for the Parkes
data, and 0.3 mJy per 40\arcsec\ beam for the combined data. The measured
value for the sensitivity of the ATCA data is in excellent agreement
with the theoretical noise estimate of 0.3 mJy per 40\arcsec\ beam for our
selected observing strategy and deconvolution scheme. Measuring the
sensitivity of the ATCA and ATCA+Parkes data is somewhat complicated
by the large number of point sources in the sky at 1.4~GHz, limiting
the size of blank regions where a noise estimate can be reliably
measured. To verify our sensitivity estimate, we produced a median
filtered version of the ATCA+Parkes image. The filtering operation,
implemented in the GIPSY routine {\tt mfilter}, moves a
2\arcmin.5$\times$2\arcmin.5 window across the map, replacing the
central pixel value ($S_{cpix}$) with the median value of the window
($S_{med}$) if $|S_{cpix}-S_{med}| > S_{med}$ + 1~mJy
\citep{vanderhulstetal92}.  The 1~mJy offset prevents unnecessary
filtering in noisy regions where the median is close to
zero. Repeating the sensitivity measurements using the median filtered
version of the ATCA+Parkes map - where it was possible to measure the
noise over much larger blank regions of sky - indicated rms values
between 0.25 and 0.35~mJy per 40\arcsec\ beam, giving us confidence in our
original sensitivity estimate for the data. The total flux density of
the median filtered map is 329~Jy, i.e. the filtering operation
removed $\sim$27\% of the total emission in the original 1.4~GHz
map. Note that the filtering operation makes no distinction between
background point sources and point sources that are intrinsic to the
LMC, so this difference in flux density should not be interpreted as
the contribution of background radio galaxies to the measured LMC flux
density. We address the contribution from background galaxies in
Section~\ref{sect:bgsrc} below.

\subsection{Total flux density of the LMC}
\label{sect:totflux}
\noindent For our final combined ATCA+Parkes map, we measure a total
flux density of 443~Jy within a 10.8\D $\times$ 12.3\D\ field centred
on (05h20m,-68d44m)$_{J2000}$. The total flux density in the
Parkes map over this same region is 413~Jy. The difference in flux
density between the final merged map and the Parkes data is due to the
relative calibration factor of 1.07 that we derived by comparing
strong point sources in the Parkes and ATCA datasets. We note that the
flux density of our Parkes 1.4~GHz map is $\sim20\%$ less than the
flux density quoted for the same map by
\citet[529$\pm$29~Jy,][]{kleinetal89}. This discrepancy can mostly be
traced to the larger beam size that we have adopted for the Parkes
data. Rather than the nominal HPBW of 15\arcmin, we determined an
effective beam width of 16.55\arcmin\ by fitting point sources in the
Parkes 1.4~GHz map with 2-dimensional Gaussians. We believe that the
effective beam width provides a more reliable estimate of the beam
size of the Parkes data, since the gridding and scanning process used
during observations and data reduction is known to slightly broaden
the HPBW \citep[e.g.][]{filipovicetal95}. Re-calculating the total
flux density in the Parkes map using the nominal beam size of
15.0\arcmin\ gives 503~Jy. We believe that the remaining discrepancy
of $\sim20$~Jy between this value and the value determined by
\citet{kleinetal89} is due to methodological differences for measuring
the LMC's integrated flux density. While we simply sum all the
emission within the rectangular 10.8\D $\times$ 12.3\D\ map area,
\citet{kleinetal89} performed an integration in elliptical rings,
including a correction for non-zero baselines. Towards the map edges,
the Parkes data exhibits a small negative offset. By blanking pixels
with negative values in the Parkes data and recalculating the
integrated flux density over the remaining unmasked area, we find that
the pixels at the map edges make an overall negative contribution of
$\sim -20$~Jy to the measured flux density. For all further discussion
in this paper, we confine our analysis to the central 7\D.5 $\times$
7\D.5 region of our map, an area that comfortably encloses the LMC.

\begin{table*}
\caption{The flux density, resolution and sensitivity of our original
  and median-filtered 1.4~GHz ATCA+Parkes maps, and the 4.8~GHz map by
  \citet{haynesetal91}. The second column lists the resolution (HPBW)
  of each image. The third column lists the flux density within the
  common 7\D.5 $\times$ 7\D.5 area surveyed at both frequencies
  (Region 1). The fourth column lists the flux density within the IRIS
  60$\mu$m 0.9 MJy~sr$^{-1}$ contour \citep[Region
    2][]{mivilledescheneslagache05}. The fifth column lists the flux
  density within the 40~mJy~beam$^{-1}$ brightness threshold of the
  4.8~GHz map, which defines the edges of our spectral index map
  (Region 3). The boundaries of these three regions are indicated in
  Fig.~\ref{fig:spdxmap}. The sixth column lists the sensitivity of
  each map, measured from blank regions of sky.  }
\label{tbl:totflux}
\begin{center}
\begin{tabular}{l|r|r|r|r|r|}
\hline
Frequency & Beam & Flux Density & Flux Density &  Flux Density &  Sensitivity \\
          &      & Region 1 & Region 2 & Region 3 & \\
(GHz) & (arcsec)  & (Jy) & (Jy) &  (Jy) & \\
\hline
1.4        &  40   &    426   &  390 &  309 & 0.3 mJy per 40\arcsec\ beam       \\
1.4-mf     &  40   &    367   &  350 &  277 & 0.3 mJy per 40\arcsec\ beam        \\
4.8        &  288  &    296   &  291 &  250 &   9 mJy per 4\arcmin.8 beam     \\
\hline
\end{tabular}
\end{center}
\end{table*}

\section{RESULTS}
\label{sect:results}

\subsection{Radio continuum morphology}
\label{sect:morphology}
\begin{figure}
\begin{center}
\includegraphics[width=12cm,angle=0]{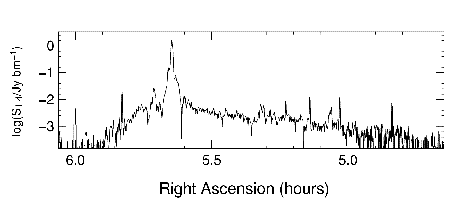}\hfill
\includegraphics[width=12cm,angle=0]{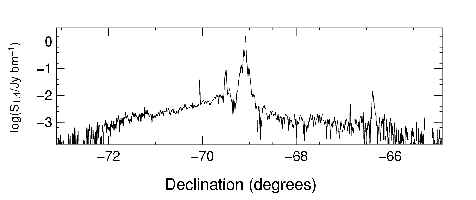}
\caption{E-W ({\it top}) and N-S ({\it bottom}) intensity
  profiles through 30~Doradus. The profiles are shown with a
  logarithmic intensity scale to emphasise the behaviour of the
  diffuse emission.}
\label{fig:profs}
\end{center}
\end{figure}

\noindent At 1.4~GHz, the radio continuum emission in the LMC is
clearly dominated by the emission associated with the 30~Doradus
region (see Figs.~\ref{fig:map} and~\ref{fig:shassa}). Across the LMC
disk, there are numerous peaks of radio emission associated with
individual \HII\ regions and supernova remnants, but the distribution of
the diffuse emission is quite asymmetric, showing a steep decline
along the eastern edge of the LMC and a more gradual decrease with
increasing distance from 30~Doradus in other
directions. Fig.~\ref{fig:profs} presents N-S and E-W intensity
profiles through 30~Doradus. As well as the abrupt eastern edge of the
LMC's radio continuum emission, the profiles reveal that the intensity
of the diffuse emission declines more slowly towards the south of
30~Doradus than to the north or west. There is minimal diffuse
emission along the western edge of the LMC, even surrounding active
star-forming complexes such as N11 and N87. We discuss a possible
explanation for the asymmetric distribution of the diffuse radio
emission in the LMC, and for the slower decline in emission's
intensity south of 30~Doradus, in Section~\ref{sect:spatialvariation}.

\subsection{Contribution from background point sources}
\label{sect:bgsrc}
\noindent Since the LMC subtends a large area on the sky, it is
possible that background radio galaxies make a non-negligible
contribution to the total 1.4~GHz flux density that we measure for the
LMC. We have attempted to estimate this contribution using a direct
and an indirect method. In the direct method, we used a
smooth-and-mask technique to measure the point source flux in sixteen 1\D
$\times$ 1\D\ control fields around the edges of our 1.4~GHz
ATCA+Parkes map. We first convolved the ATCA+Parkes data with a
1\arcmin\ Gaussian kernel, and masked all pixels in the 4 control
fields where the flux density of the smoothed map was less than
0.8~mJy~beam$^{-1}$. We measured the total flux density of the unmasked
pixels in each of the control fields, and then calculated the mean
flux density per square degree. Multiplying this by the area of the
LMC yields an estimate of the flux density due to background
sources. Assuming that the LMC has an angular size of 7.5\D $\times$
7.5\D, this method indicates that the 1.4~GHz flux density of sources
behind the LMC is $46\pm10$~Jy, approximately 10\% of the LMC's measured
total flux density. \\

\noindent We made a second, indirect estimate of the flux density due
to background sources using the differential 1.4~GHz source count
distributions of the FIRST and NVSS surveys determined by
\citet{blakewall02}. We fitted a curve to the data presented in their
fig.~5, and calculated the total flux density expected for sources
in the flux density range [0.001,1]~Jy within an area of 7.5\D
$\times$ 7.5\D. The predicted 1.4~GHz flux density of background
sources is $52\pm3$~Jy from the FIRST source count distribution,
and $55\pm3$~Jy from the NVSS data. These values are in
reasonable agreement with our direct estimate from the ATCA+Parkes
map, given the potential sources of uncertainty in our direct
estimate.\\

\subsection{Thermal fraction of radio emission at 1.4~GHz}
\label{sect:thermalfraction}
\begin{figure}
\begin{center}
\includegraphics[width=15cm,angle=0]{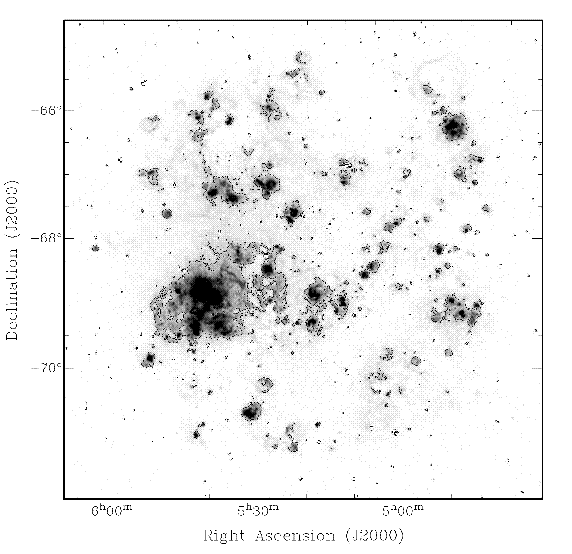}
\caption{SHASSA map of \Ha\ emission from the LMC
\citep{gaustadetal01}. The black contours indicate an \Ha\ brightness
of 50~R. A square-root intensity scale has been used to emphasise the
characteristics of the diffuse emission. Note that while the bright
features in the \Ha\ and 1.4~GHz radio continuum maps show good
correspondence, the diffuse emission is brighter and more extended at
radio wavelengths, particularly in the south-east of the LMC and in a
central region around (5h20m,-68d30m)$_{J2000}$ (compare Fig.~\ref{fig:map}).}
\label{fig:shassa}
\end{center}
\end{figure}

\noindent Many of the brightest features in the 1.4~GHz radio map are
also observed to be bright sources of \Ha\ emission (see
Fig.~\ref{fig:shassa}). This suggest that a significant fraction of
the total 1.4~GHz radio continuum flux density may be produced by a
few very bright star-forming regions, e.g. 30~Doradus, N11 and N44. We
used the Southern \Ha\ Sky Survey Atlas (SHASSA) map of \Ha\ emission
in the LMC to obtain a rough estimate for the fraction of the LMC's
1.4~GHz flux density that is of thermal origin
\citep{gaustadetal01}.\footnote{The Southern \Ha\ Sky Survey Atlas
  (SHASSA) is supported by the National Science Foundation.}  Our
1.4~GHz ATCA+Parkes map and the SHASSA map were regridded to a common
pixel scale of 20\arcsec, and a common 7.5\D $\times$ 7.5\D field of
view. We measured the total flux density in the 1.4~GHz map for pixels
with \Ha\ emission above brightness thresholds of 100 and 500~R. The
1.4~GHz flux densities corresponding to these \Ha\ brightness
thresholds are 140 and 81~Jy, suggesting that the thermal fraction is
of the LMC's 1.4~GHz radio continuum emission is likely to be greater
than 20\%. Note that this estimate should be understood as providing
only a general indication that the thermal fraction of the LMC's radio
continuum emission is higher than for normal spiral galaxies, a
phenomenon that appears to be relatively common amongst dwarf galaxies
\citep[e.g.][]{kleinetal84}. We have not accounted for extinction of
the \Ha\ emission, nor for the fact that most evolved \HII\ regions
will contain a mixture of nonthermal and thermal radio emission.  We
note that the flux density contained within a 0\D.5 $\times$ 0\D.5 box
centred on 30~Doradus is 56~Jy, suggesting that 30~Doradus alone
accounts for a significant fraction of the LMC's total and thermal
radio flux density.

\subsection{Star formation in the LMC}
\label{sect:sfr}
\noindent Since radio emission is unaffected by extinction, developing
a reliable calibration for a galaxy's star formation rate (SFR) based
on its integrated radio flux density would be a valuable tool for
studies of star formation at high redshift. However, the calibrations
derived from local samples typically reflect the characteristic
thermal/nonthermal emission ratio observed for normal galaxies
\citep[$\sim$10\%, e.g.][]{condon92}, and generally do not hold for
young starbursts or dwarf galaxies
\citep[e.g.][]{rousseletal03,huntbianchimaiolino05}. Recently, \citet{bell03}
compared radio-derived SFRs to measurements determined from
far-ultraviolet, infrared (IR) and optical data for a sample of 249
normal and irregular galaxies, and presented new radio and IR SFR
calibrations that take account of the low dust opacity and lower
fraction of nonthermal radio emission in dwarf galaxies. We used
Equation 6 from \citet{bell03} to estimate the LMC's star formation
rate from its total radio continuum flux density, finding SFR$_{RC}
\sim$0.2 \msol\ yr$^{-1}$. This value agrees very well with the SFRs
derived from the LMC's total infrared (TIR) and \Ha\ luminosities. To
estimate SFR$_{TIR}$, we used the SED models presented in
\citet{dalehelou02} to translate the Improved Reprocessing of the IRAS
Survey (IRIS) 60- and 100$\mu$m integrated fluxes of the LMC into an
estimate of the TIR flux, and Equation 5 from \citet{bell03} to
calculate the SFR$_{TIR}$, finding SFR$_{TIR} \sim$0.2
\msol\
yr$^{-1}$.\footnote{http://www.cita.utoronto.ca/$\sim$mamd/IRIS/} The
LMC's total \Ha\ luminosity, corrected for an average extinction of
$A_{V}$ = 1 across the LMC, is $4.1 \times 10^{41}$ erg s$^{-1}$
\citep{kennicutthodge86}. Applying the SFR calibration for
\Ha\ emission recently published by \citet{calzettietal07} also
results in SFR$_{H\alpha} \sim$0.2 \msol\ yr$^{-1}$. While the SFRs
derived using these three tracers are in excellent agreement, the
supernova rate that they predict is $\sim0.0016$ yr$^{-1}$, assuming a
Salpeter initial mass function between 0.1 and 100~\msol. If the LMC
followed the standard relation observed for normal spiral galaxies
\citep{condon92}, this supernova rate should generate a nonthermal
radio flux density of $\sim$530~Jy at 1.4~GHz, $\sim$20\% greater than
the LMC's total observed 1.4~GHz flux density. It thus appears that
while the LMC follows the radio-FIR correlation for normal galaxies
\citep{hughesetal06}, it does not conform to the usual scaling
relations between star formation and radio emission.

\subsection{The radio spectral index}
\label{sect:spectralindex}
\noindent We used the Parkes map of the LMC at 4.8~GHz published by
\citet{haynesetal91} in combination with our 1.4~GHz map to verify the
LMC's total radio spectral index, and to investigate spatial
variations of the radio spectral index within the LMC. We calculated
the global spectral index using i) the total flux densities measured
over the commonly observed area at both frequencies, ii) the total
flux densities measured for a region where the 60$\mu$m emission in
the IRIS map of the LMC is greater than 0.9~MJy~sr$^{-1}$
\citep{mivilledescheneslagache05}, and iii) the total flux densities
measured over a smaller region where the 4.8~GHz emission was brighter
than 40~mJy~beam$^{-1}$. The flux density of the LMC at 1.4 and 4.8~GHz
within each of these three regions is listed in
Table~\ref{tbl:totflux}. The boundaries of each region are indicated
on the map of the LMC in Fig.~\ref{fig:spdxmap}.\\

\noindent Our three estimates for the LMC's global radio spectral
index are quite similar, ranging from $\alpha=-0.29$ for the
measurement across the common 7.5\D $\times$ 7.5\D\ field, to
$\alpha=-0.17$ for the more restricted region where the 4.8~GHz
emission is brighter than 40~mJy~beam$^{-1}$. All three estimates are
much flatter than the typical spectral index of normal spiral galaxies
at these frequencies \citep[$\alpha \sim
  -0.74\pm0.12$,][]{gioiaetal82}, suggesting that the thermal fraction
of the LMC's radio continuum emission at 1.4~GHz is indeed relatively
large. We note that our estimate is flatter than the spectral index
between 20~MHz and 2.3~GHz determined for the LMC by \citet[$\alpha =
  -0.56\pm0.05$,][]{kleinetal89}, but consistent with the spectral
index at higher frequencies, $\alpha = -0.3\pm0.1$, indicated by the
\citet{haynesetal91} data. \\

\subsection{Spatial variation of the radial spectral index}
\label{sect:spatialvariation}
\begin{figure}
\begin{center}
\includegraphics[width=15cm,angle=0]{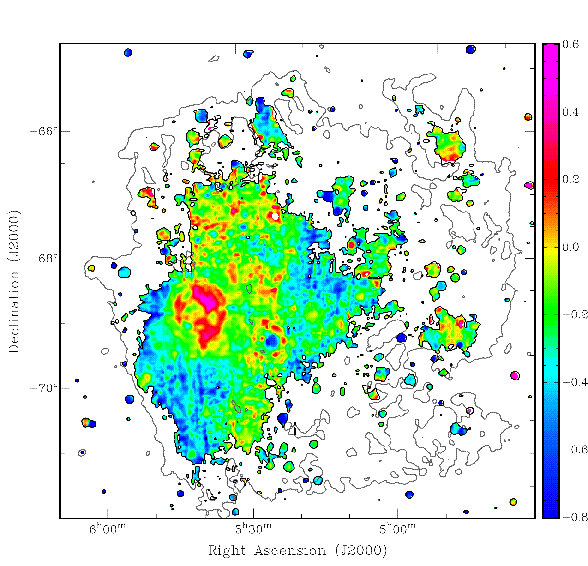}
\caption{The spectral index map calculated from a smoothed version of
  the 1.4~GHz ATCA+Parkes map and the 4.8~GHz Parkes map of
  \citet{haynesetal91}. The black line indicates the
  40~mJy~beam$^{-1}$ brightness threshold of the 4.8~GHz map. The grey
  line reproduces the 0.9~MJy~sr$^{-1}$ contour of the IRIS $60\mu$m
  map of \citet{mivilledescheneslagache05}, which approximately marks
  the boundary of the LMC's disk. }
\label{fig:spdxmap}
\end{center}
\end{figure}

\noindent To investigate spatial variations of the radio spectral
index, we also produced a spectral index map. The map was constructed
by smoothing the ATCA+Parkes 1.4~GHz image to the resolution of the
Parkes data (4\arcmin.8), blanking pixels below a certain brightness
threshold, and calculating the spectral index of the remaining high
signal-to-noise pixels. For both maps, we blanked pixels with a flux
density less than 40~mJy~beam$^{-1}$ in the 4.8~GHz data; it was not
necessary to perform additional masking based on the 1.4~GHz map since
the 1.4~GHz data are more sensitive than the Parkes 4.8~GHz data. The
resulting spectral index map is shown in Fig.~\ref{fig:spdxmap}. We
note that our masking technique is biased against faint emission with
a steep nonthermal spectral index, since pixels lacking high
signal-to-noise 4.8~GHz emission are excluded, even if 1.4~GHz
emission at that position is well-detected. The significance of this
effect can be gauged by comparing our estimates of the global spectral
index calculated over the three different fields of view. The common
7.5\D $\times$ 7.5\D\ field encloses approximately four times the
number of unmasked pixels as our spectral index map, but the global
radio spectral index determined for these two regions only varies by
$\Delta \alpha \sim 0.1$. This suggests that the average spectral
index in regions of the LMC excluded by the 40~mJy~beam$^{-1}$
brightness threshold in the 4.8~GHz map is only slightly steeper than
for the regions that are included in our spectral index map.\\

\noindent The dominant feature of the spectral index map is the
emission associated with 30~Doradus. The spectral index appears to
become more negative with increasing distance from 30~Doradus,
suggesting an increasing nonthermal fraction at larger radii. This
transition is asymmetric however: to the south-east of 30~Doradus, the
spectral index decreases much more abruptly than in the north-west
direction. Several of the LMC's well-known star-forming regions can
also be identified in the spectral index map (e.g. N11, N87), and
these exhibit a variegated pattern of positive and negative spectral
indices. Overall, spectral indices across the LMC are relatively flat,
and very few regions have spectral indices more negative than
$\alpha\sim-0.7$. The most negative spectral indices, moreover, are
associated with compact sources that are probably SNRs or background
radio galaxies. A notable exception is an extended region of radio
continuum emission with $\alpha\sim-0.4$ located at the interface of
the LMC4 and LMC5 superbubbles, at an approximate position of (05h25m
-66d15m)$_{J2000}$. The \hi, radio continuum and 8.3$\mu$m emission
from this region were studied in detail by \citet{cohenetal03}, who
identified the region as a site of triggered secondary star formation,
containing a mixture of young massive stars and recently exploded
supernova remnants. \\

\begin{figure}
\begin{center}
\includegraphics[width=12cm]{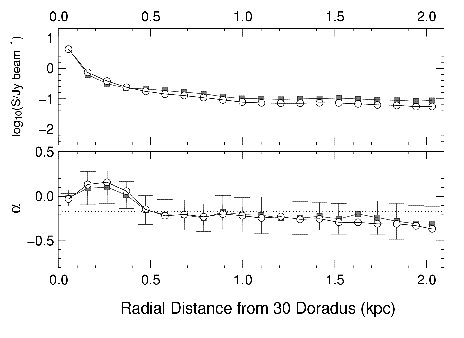}
\caption{Variation in the radio flux density and spectral index,
  $\alpha$, as a function of distance from 30~Doradus. The top panel
  shows the radial variation in the 1.4 (grey squares) and 4.8~GHz
  (open circles) flux density. The bottom panel shows the variation of
  the spectral index, as determined from the spectral index map (grey
  squares) and derived from the average 1.4 and 4.8~GHz flux density
  in each annulus (open circles). The error bars indicate the standard
  deviation of the pixel values in each annulus. The horizontal dotted
  line indicates the LMC's global spectral index that we measure,
  $\alpha=-0.23$.}
\label{fig:spdxcmp}
\end{center}
\end{figure}

\noindent In order to quantify the behaviour of the spectral index
with increasing distance from 30~Doradus, we measured the average
spectral index in circular annuli centred on
(05h38m42s,-69d06m03s)$_{J2000}$. The width of each annulus was
7\arcmin.2 to ensure that our measurements of the spectral index were
statistically independent. We determined the average spectral index
using two methods: i) using the spectral index map shown in
Fig.~\ref{fig:spdxmap} directly, and ii) calculating the
spectral index from the average 1.4~GHz and 4.8~GHz flux density in
each annulus. The average spectral indices determined by the two
methods are in good agreement. A plot showing the behaviour of the
spectral index as a function of distance from 30~Doradus is shown in
Fig.~\ref{fig:spdxcmp}.\\

\noindent The plots in Fig.~\ref{fig:spdxcmp} show that the radio
spectral index departs from the average LMC value close to
30~Doradus. Within $R\sim0.5$~kpc of 30~Doradus, the spectral index is
more positive than the average value for the LMC, showing a pronounced
bump that increases to positive values ($\alpha \sim 0.1$) at
$R\sim0.2$~kpc and then falls back to $\alpha \sim -0.2$ at
$R\sim0.5$~kpc. These positive spectral index values are visible in
the spectral index map, particularly to the north-west of
30~Doradus. Clearly a high fraction of the radio continuum emission
close to 30~Doradus is of thermal origin, but it is not clear why the
spectral index should increase and then decline over these spatial
scales. Positive values of the radio spectral index can indicate
regions where there is signicant free-free absorption, which have been
identified with extremely young ($\leq 1$~Myr), dense, heavily
embedded star clusters \citep[e.g.][]{johnsonkobulnicky03}. However
these ``ultradense \HII\ regions'' are typically very compact ($R\sim2$
to 4~pc) and should be associated with extremely dense molecular gas
\citep{elmegreen02}: neither condition appears to be satisfied
here. Based on a spectral analysis of OB stars in 30~Doradus,
\citet{walbornblades97} have suggested that the most recent star
formation activity in 30~Doradus is occurring to the north-west of the
central R136 cluster, but their observations are again on much smaller
spatial scales than those indicated here. At present, we have no good
explanation for this feature. We note, however, that the feature is
also present in a spectral index map constructed from Parkes data
alone, which seems to rule out the possibility that it is an artefact
of the ATCA 1.4~GHz data.\\

\noindent Beyond $R\sim0.5$~kpc, the average spectral index exhibits a
shallow decline with increasing distance from 30~Doradus, although the
dispersion within each annulus is quite large. This steepening of the
spectral index with increasing radius might indicate that the
30~Doradus region is the primary site of cosmic ray electron
production in the LMC, and that synchrotron and inverse Compton losses
are occurring as the relativistic electrons propagate away from the
central star-forming region. Inspection of the spectral index map
shows that this steepening of the spectral index is somewhat
asymmetric: the spectral index remains relatively flat ($\alpha \sim
-0.1$) to the north and west of 30~Doradus, but decreases more sharply
towards the south and east to values approaching $\alpha \sim
-0.4$. The south-eastern region of the LMC is characterised by very
high \hi\ column densities \citep[e.g.][]{staveleysmithetal03}, and
radio continuum emission that is strongly polarised
\citep[e.g.][]{kleinetal93,gaensleretal05b}. The NANTEN survey of
$^{12}$CO emission in the LMC shows that there is a long filament of
molecular gas at this location, but the star-forming activity of this
massive molecular cloud is relatively low
\citep{fukuietal01,kawamuraetal05}. A plausible physical scenario to
explain both the relatively strong diffuse radio emission and the
rapid steepening of the spectral index in this region would be an
increased magnetic energy density. For a population of cosmic rays
with relativistic energy spectrum $dN/dE \propto E^{-2}$,
corresponding to a radio spectral index of $\alpha=-0.5$ at 1~GHz, the
synchrotron emissivity scales as $B^{1.5}$. An amplified magnetic
field, combined with the high gas densities and the intense radiation
field surrounding 30~Doradus, would shorten the cooling time-scale of
the electrons in this region, causing the radio spectral index to
steepen. A stronger magnetic field is consistent with the unusually
low star-forming activity of the molecular clouds in this region,
since higher magnetic pressures should also impede cloud collapse. \\

\section{SUMMARY AND FINAL REMARKS}
\label{sect:summary}
\noindent We present a sensitive ATCA+Parkes mosaic image of the 1.4~GHz radio
continuum emission from the LMC that is suitable for studying emission
on all scales greater than 40\arcsec. The reduction of this data
involved a two-step deconvolution procedure that is able to recover
both extended and point-like emission features, and a peeling
technique that succesfully removed calibration errors from bright
off-axis sources. Our analysis of this data has shown: \\

\noindent 1. The diffuse radio continuum emission in the LMC has an asymmetric
morphology, showing a steep decline along the eastern edge of the LMC
and a more gradual decrease with increasing distance from 30~Doradus
elsewhere. The intensity of the diffuse emission surrounding
30~Doradus remains stronger towards the south than in other
directions, and is apparently correlated with the high \hi\ column
densities in this region. A similar asymmetric morphology is seen in
the 1.4-4.8~GHz spectral index map of the LMC. We suggest that a
plausible explanation for these features is the amplification of the
magnetic field in this region, perhaps due to field compression
resulting from the motion of the LMC through the Milky Way's halo.\\

\noindent 2. The total radio flux density of the LMC at 1.4~GHz is 426~Jy, of
which $\geq$20\% probably has a thermal origin. The star formation
rate implied by the LMC's total 1.4~GHz emission is 0.2
\msol\ yr$^{-1}$, in good agreement with star formation rates derived
from its infrared and \Ha\ luminosity
\citep{bell03,calzettietal07}. This level of star formation
significantly overpredicts the nonthermal flux density that should be
generated by supernovae however, suggesting that the LMC does not
exhibit the same relationship between star formation and radio
emission as normal spiral galaxies \citep[e.g][]{condon92}.\\

\noindent Finally, we note that these radio continuum data are
available for use by the wider astronomical community. Our group is
currently undertaking a study of the LMC's magnetic field, preparing a
catalogue of 1.4~GHz point sources in the LMC, and comparing the LMC's
1.4~GHz radio continuum and far-infrared dust emission. However, the
data should also be very useful for studies of individual objects such
as supernova remnants, \HII\ regions and planetary nebulae, and for
detailed, galaxy-wide comparisons between the radio continuum emission
and other tracers of star formation. The data may be retrieved from
the website http://www.atnf.csiro.au/research/lmc\_ctm/. \\

\section{ACKNOWLEDGMENTS}
\label{sect:thanks}
\noindent AH would like to thank Tim Cornwall, Urvashi Rau and Enno
Middelberg for useful discussions regarding peeling and synthesis
imaging techniques. We also thank John Dickel, Marc-Antoine
Miville-Desch\^{e}nes and Guilaine Lagache for access to the Parkes
4.8~GHz and IRIS 60 and 100~$\mu$m images that were used in this
paper. \\

\bibliographystyle{mn}
\label{sect:bibliography}
\bibliography{radiofir,lmc,obstools,mnemonic,software,sf,rc,ir}
\label{lastpage}

\end{document}